\documentclass[twocolumn,amsmath,aps,fleqn]{revtex4}
\usepackage{graphicx,amssymb}
\begin{document}
\newcommand{\be}{\begin{equation}}
\newcommand{\ee}{\end{equation}}
\newcommand{\bq}{\begin{eqnarray}}
\newcommand{\eq}{\end{eqnarray}}
\newcommand{\bsq}{\begin{subequations}}
\newcommand{\esq}{\end{subequations}}
\newcommand{\bc}{\begin{center}}
\newcommand{\ec}{\end{center}}
\newcommand {\R}{{\mathcal R}}
\newcommand{\al}{\alpha}
\newcommand\lsim{\mathrel{\rlap{\lower4pt\hbox{\hskip1pt$\sim$}}
    \raise1pt\hbox{$<$}}}
\newcommand\gsim{\mathrel{\rlap{\lower4pt\hbox{\hskip1pt$\sim$}}
    \raise1pt\hbox{$>$}}}

\title{Eddington-inspired Born-Infeld gravity: nuclear physics constraints and the validity of the continuous fluid approximation}

\author{P.P. Avelino}
\email[Electronic address: ]{ppavelin@fc.up.pt}
\affiliation{Centro de Astrof\'{\i}sica da Universidade do Porto, Rua das Estrelas, 4150-762 Porto, Portugal}
\affiliation{Departamento de F\'{\i}sica e Astronomia, Faculdade de Ci\^encias, Universidade do Porto, Rua do Campo Alegre 687, 4169-007 Porto, Portugal}

\date{\today}
\begin{abstract}

In this paper we investigate the classical non-relativistic limit of the Eddington-inspired Born-Infeld theory of gravity. We show that strong bounds on the value of the only additional parameter of the theory $\kappa$, with respect to general relativity, may be obtained by requiring that gravity plays a subdominant role compared to electromagnetic interactions inside atomic nuclei. We also discuss the validity of the continuous fluid approximation used in this and other astrophysical and cosmological studies. We argue that although the continuous fluid approximation is expected to be valid in the case of sufficiently smooth density distributions, its use should eventually be validated at a quantum level.

\end{abstract}
\maketitle

\section{\label{intr}Introduction}

A key property of the Eddington-inspired Born-Infeld (EiBI) theory of gravity introduced in \cite{Banados:2010ix} is that it is equivalent to Einstein's general relativity in vacuum. This theory has many additional attractive features: it may lead to a non-singular description of the Universe \cite{Banados:2010ix} (see also \cite{EscamillaRivera:2012vz,Liu:2012rc,Avelino:2012ue}) and it may accommodate compact stars made of pressureless dust \cite{Pani:2011mg,Avelino:2012ge,Pani:2012qb} (see also \cite{Deser:1998rj,Vollick:2003qp,Wohlfarth:2003ss,Nieto:2004qj,Comelli:2005tn,Vollick:2005gc,Zinoviev:2006,Ferraro:2008ey,Fiorini:2009ux,Ferraro:2009zk,Gullu:2010pc,Alishahiha:2010iq,Gullu:2010em,DeFelice:2012hq} for other relevant studies of Born-Infeld type gravitational models and \cite{Clifton:2011jh} for a recent review on alternative theories of gravity).

Astrophysical constraints on the single extra parameter of the theory $\kappa$ (with respect to General Relativity) have been determined considering the physics of astronomical objects such as neutron stars \cite{Pani:2011mg,Avelino:2012ge,Pani:2012qb}, the Sun \cite{Casanellas:2011kf,Avelino:2012ge} or the Universe as a whole \cite{Avelino:2012ge}, with the tightest constraints being associated to smaller objects/earlier cosmological times.  With the fundamental parameters of the theory (the gravitational constant $G$, the speed of light in vacuum $c$ and $\kappa$) it is possible to define a fundamental length, mass and density respectively by $l_f=|\kappa|^{1/2} G^{-1/2}$, $m_f=|\kappa|^{1/2}c^2G^{-3/2}$ and $\rho_f=c^2|\kappa|^{-1}$. In \cite{Avelino:2012ge} it has been shown that for $\kappa>0$ an effective pressure arises which may prevent gravitational collapse of pressureless matter (see also \cite{Delsate:2012ky}). It has been further demonstrated that $l_f$ and $\rho_f$ determine the minimum radius and maximum density that a stable non-singular compact object held together by gravity might have.

In this letter we consider the classical non-relativistic limit of EiBI gravity. In Sec. II we generalize the classical analysis of \cite{Avelino:2012ge} to account for the contribution of an electric field and we estimate, as a function of $\kappa$, the scale below which gravity becomes more important than the electrostatic interaction. In Sec. III these results are used to obtain an upper limit on the value of  $|\kappa|$, assuming that gravity plays a subdominant role compared to the electromagnetic interaction inside nuclei. In Sec. IV we critically assess the continuous fluid approximation used in this paper and in previous astrophysical and cosmological studies. We conclude in Sec. V.

\section{EiBI gravity versus electrostatic interaction}

In \cite{Banados:2010ix} it was shown that, in the non-relativistic limit, the EiBI theory of gravity leads to a modified Poisson equation given by
\be
\nabla^2 \phi = 4 \pi G \rho_m+\frac{\kappa}{4}\nabla^2 \rho_m\,,
\label{poisson}
\ee
where $\phi$ is the gravitational potential and $\rho_m$ is the matter density. The gravitational acceleration generated by the matter density distribution is given by
\be
{\bf a}_m=-\nabla \phi\,,
\ee
so that
\be
\nabla \cdot {\bf a}_m=-4\pi G \rho_m-\frac{\kappa}{4}\nabla^2 \rho_m\,.
\ee

The electric field generated by a spherically symmetric distribution of charge can be calculated using
\be
\nabla \cdot {\bf E}=\frac{\rho_q}{\epsilon_0}=\frac{q}{m} \frac{\rho_m}{\epsilon_0} \,,
\ee
where $\epsilon_0$ is the vacuum permittivity and $q/m$ is the charge to mass ratio. The acceleration due to the electric field is given by
\be
{\bf a}_q=\frac{q}{m} {\bf E}\,,
\ee
so that
\be
\nabla \cdot {\bf a}_q=\left(\frac{q}{m}\right)^2 \frac{\rho_m}{\epsilon_0}\label{divaq} \,.
\ee
For simplicity, we have assumed a constant charge to mass ratio $q/m$ in the derivation of Eq. (\ref{divaq}).

The acceleration of the particles due to the gravitational and electric fields is
\be
{\bf a}={\bf a}_m+{\bf a}_q\,,
\ee
whose divergence is given by
\be
\nabla \cdot {\bf a}=-\left(\frac{\kappa}{4}\nabla^2+4\pi G \beta\right)\rho_m\,.
\ee
where
\be
\beta=1-\frac{1}{4 \pi G \epsilon_0}\left(\frac{q}{m}\right)^2\,.
\ee
Moving to Fourier space one obtains
\be
i{\bf k} \cdot {\bf a}_{\bf k}=\left(k^2\frac{\kappa}{4}-4\pi G \beta\right)\rho_{m{\bf k}}\,.
\ee
If $q/m=0$ then $\beta=1$. In this case, and assuming that $\kappa>0$, the condition ${\bf k} \cdot {\bf a}_{\bf k}=0$ defines the Jeans wavenumber and wavelength by
\be
k_J={\sqrt {\frac{16\pi G}{\kappa}}}\,, \qquad \lambda_J=\frac{2\pi}{k_J}={\sqrt \frac{\pi \kappa}{4 G}}\,,
\ee
respectively. Here, $\lambda_J$ is independent of the matter density, being approximately equal to the fundamental scale of the theory $l_f$. The effective Jeans length defines the critical scale below which the collapse of pressureless dust is not possible, determining  the minimum scale of compact objects which are held together by gravity. 

On the other hand, if $\beta < 0$ then the condition ${\bf k} \cdot {\bf a}_{\bf k}=0$ is no longer possible. Still, one may calculate the scale for which 
\be
|{\bf k} \cdot {\bf a}_{m{\bf k}}|=|{\bf k} \cdot {\bf a}_{q{\bf k}}|\,. \label{gqeq}
\ee
If $|\beta| \gg 1$ then Eq. (\ref{gqeq}) defines the scale
\be
\lambda_g =\pi \frac{m}{q} {\sqrt {\epsilon_0 |\kappa|}} \label{lambdag}\,,
\ee
bellow which EiBI gravity is expected to dominate over the electromagnetic interaction.   

\section{Nuclear constraints}

The classical electron radius
\be
r_e=\frac{e^2}{4 \pi \epsilon_0 m_e c^2} \sim 3 \times 10^{-15} \, {\rm m}\,,
\ee
may be obtained (up to a constant of order unity) by requiring the electrostatic potential energy $U_e$ of a uniform spherical distribution of charge of radius $r_e$ and total charge $e$ to be equal to the relativistic energy of the electron
\be
U_e \sim \frac{e^2}{4\pi \epsilon_0 r_e}=m_e c^2\,,
\ee
where we have ignored the factor of $3/5$ in $U_e$.
The condition $\lambda_g < r_e$, with $q/m=e/m_e$, implies that
\be
|\kappa| < \frac{e^6}{16 \pi^4 \epsilon_0^3 m_e^4 c^4} = 3 \times 10^3 \, {\rm kg^{-1} \, m^5 \, s^{-2}}\,.
\label{constraint1}
\ee
Here, $e$ and $m_e$ are the electron charge and mass, respectively.

On the other hand, the radius of atomic nuclei lie in the range
 \be
r_n \sim 10^{-15}-10^{-14}\, {\rm m}\,.
\ee
while the charge to mass ratio squared is smaller in atomic nuclei than that of  the electron by a factor of $(m_e/m_p)^2 \sim 3 \times 10^{-7}$. We shall assume that gravity plays a negligible role in radioactive alpha decay, whereby an atomic nucleus spontaneously emits a helium nucleus (alpha particle). This implies that $\lambda_g <  r_n$, thus leading to
\be
|\kappa| < 10^{-3} \, {\rm kg^{-1} \, m^5 \, s^{-2}}\,,
\ee
a constraint which is stronger by about $7$ orders of magnitude than that given in Eq. (\ref{constraint1}). Coincidentally, this constraint is of similar magnitude (albeit slightly stronger) to that obtained using neutron stars. However, it should be regarded as an order of magnitude constraint since it is based on a purely classical non-relativistic analysis, relying on the validity of continuous fluid approximation which we shall discuss in the following section.

\section{Validity of the continuous fluid approximation}

The continuous fluid approximation has been used in all EiBI gravity studies performed so far. However, we shall show that it might not always be valid. By defining an auxiliary field 
\be
\phi'=\phi-\kappa \rho_m/4\,,
\ee
the gravitational acceleration can be written as
\be
{\bf a}_m=-\nabla \phi' +  \delta {\bf a}_m\,,
\ee
where $\phi'$ satisfies the standard Poisson equation
\be
\nabla^2 \phi'=4\pi G \rho_m\,,
\ee
and
\be
 \delta {\bf a}_m=-\frac{\kappa}{4}\nabla \rho_m\,,
 \label{pecacc}
\ee
is the peculiar acceleration in EiBI gravity.
Consider a classical object whose mass density is a continuous function of ${\bf r}= (x,y,z)$ ($x$, $y$ and $z$ are cartesian coordinates). If $\rho_m({\bf r})=0$ for $x > x_{max}$ and $x < x_{min}$ then
\be
\int_{x_{min}}^{x_{max}} dx  \rho_m \frac{\partial \rho_m}{\partial x}=\left[\frac{\rho_m^2}{2}\right]_{x_{min}}^{x_{max}}=0\,,
\ee
and a similar result applies if $x$ is substituted by $y$ or $z$. This implies that $\delta {\bf a}_m$ cannot be associated to the motion of the center of mass or rotations of compact objects since
\be
\delta {\bf F} =\int_V d^3 r  \rho_m \delta {\bf a}_m={\bf 0}\,,
\ee
and
\be
\delta {\bf M} =\int_V d^3 r \rho_m {\bf r} \times \delta {\bf a}_m={\bf 0}\,,
\ee
where $\int_V$ represents an integral over the whole volume of the object. 

In the case of point particles $\rho_m({\bf r})=\sum_i m_i\delta^3({\bf r}-{\bf r}_{i})$ where $\delta^3({\bf r})$ is the three-dimensional Dirac delta function and ${\bf r}_{i}$ represents the position of the point particles of mass $m_i$. In the fluid approximation $\rho_m$ is approximated by
\be
\rho_m({\bf r})=\sum_i m_i W(|{\bf r}-{\bf r}_i|/R)\,,
\ee
where the function $W$ is a filter which produces a smooth density field on scales smaller than $R$. In the case of a top-hat filter 
\bq
W(q)&=& \frac{3}{4\pi R^3}\,, \qquad  0 \le q \le1\,,\nonumber \\
W(q)&=&0\,, \qquad \qquad  \qquad q> 1\,.
\eq

The filtering process transforms a discrete distribution of particles into a continuous energy density distribution potentially leading to unwanted fictitious interactions. Consider a statistically uniform distribution of static point masses with planar symmetry in the region defined by $|x|<a$ and let us apply the top-hat filter, transforming the discrete distribution of mass into a continuous one.  Assume that both the average inter-particle spacing $d$ and the thickness of the planar mass distribution $2a$ are much smaller than $R$. The density distribution after smoothing with a top-hat filter is such that $\rho_m(x) \propto R^2-x^2$ for $|x| \le R$ and $\rho(x)=0$ for $|x| > R$. According to Eq. (\ref{pecacc}) this gives rise to a peculiar acceleration for $|x| \le R$, which did not exist in the original mass distribution, given by
\be
\delta {\bf a}_m=-\frac{\kappa}{4} \frac{d\rho_m}{dx} {\bf e}_x \propto x {\bf e}_x \,,
\label{pecacc1}
\ee
where ${\bf e}_x$ is the unit vector in the positive $x$-direction. This example shows that, as anticipated, the filtering process may generate artificial interactions between the particles. Although this might not be a problem in the case of sufficiently smooth density distributions, the performance of the fluid approximation in EiBI theories of gravity should be further investigated in more detailed studies which consider how to bridge the quantum and classical limits of the theory.

\section{\label{conc}Conclusions}

In this paper we derived strong constraints on the Eddington-inspired Born-Infeld theory of gravity assuming that gravity plays a subdominant role compared to electromagnetic interactions inside atomic nuclei. We have also shown that  the continuous fluid approximation, used in all astrophysical and cosmological studies of EiBI gravity performed so far, should be applied with some care, as it may lead to artificial interactions between compact massive objects.  We further argued that although the continuous fluid approximation is expected to be valid in the case of sufficiently smooth density distributions, its use should eventually be validated at a quantum level.

\bibliography{EiBI}

\begin{thebibliography}{24}
\expandafter\ifx\csname natexlab\endcsname\relax\def\natexlab#1{#1}\fi
\expandafter\ifx\csname bibnamefont\endcsname\relax
  \def\bibnamefont#1{#1}\fi
\expandafter\ifx\csname bibfnamefont\endcsname\relax
  \def\bibfnamefont#1{#1}\fi
\expandafter\ifx\csname citenamefont\endcsname\relax
  \def\citenamefont#1{#1}\fi
\expandafter\ifx\csname url\endcsname\relax
  \def\url#1{\texttt{#1}}\fi
\expandafter\ifx\csname urlprefix\endcsname\relax\def\urlprefix{URL }\fi
\providecommand{\bibinfo}[2]{#2}
\providecommand{\eprint}[2][]{\url{#2}}

\bibitem[{\citenamefont{Banados and Ferreira}(2010)}]{Banados:2010ix}
\bibinfo{author}{\bibfnamefont{M.}~\bibnamefont{Banados}} \bibnamefont{and}
  \bibinfo{author}{\bibfnamefont{P.~G.} \bibnamefont{Ferreira}},
  \bibinfo{journal}{Phys. Rev. Lett.} \textbf{\bibinfo{volume}{105}},
  \bibinfo{pages}{011101} (\bibinfo{year}{2010}).

\bibitem[{\citenamefont{Escamilla-Rivera
  et~al.}(2012)\citenamefont{Escamilla-Rivera, Banados, and
  Ferreira}}]{EscamillaRivera:2012vz}
\bibinfo{author}{\bibfnamefont{C.}~\bibnamefont{Escamilla-Rivera}},
  \bibinfo{author}{\bibfnamefont{M.}~\bibnamefont{Banados}}, \bibnamefont{and}
  \bibinfo{author}{\bibfnamefont{P.~G.} \bibnamefont{Ferreira}},
  \bibinfo{journal}{Phys. Rev.} \textbf{\bibinfo{volume}{D85}},
  \bibinfo{pages}{087302} (\bibinfo{year}{2012}).

\bibitem[{\citenamefont{Liu et~al.}(2012)\citenamefont{Liu, Yang, Guo, and
  Zhong}}]{Liu:2012rc}
\bibinfo{author}{\bibfnamefont{Y.-X.} \bibnamefont{Liu}},
  \bibinfo{author}{\bibfnamefont{K.}~\bibnamefont{Yang}},
  \bibinfo{author}{\bibfnamefont{H.}~\bibnamefont{Guo}}, \bibnamefont{and}
  \bibinfo{author}{\bibfnamefont{Y.}~\bibnamefont{Zhong}},
  \bibinfo{journal}{Phys. Rev.} \textbf{\bibinfo{volume}{D85}},
  \bibinfo{pages}{124053} (\bibinfo{year}{2012}).

\bibitem[{\citenamefont{Avelino and Ferreira}(2012)}]{Avelino:2012ue}
\bibinfo{author}{\bibfnamefont{P.~P.} \bibnamefont{Avelino}} \bibnamefont{and}
  \bibinfo{author}{\bibfnamefont{R.~Z.} \bibnamefont{Ferreira}}
  (\bibinfo{year}{2012}), \eprint{1205.6676}.

\bibitem[{\citenamefont{Pani et~al.}(2011)\citenamefont{Pani, Cardoso, and
  Delsate}}]{Pani:2011mg}
\bibinfo{author}{\bibfnamefont{P.}~\bibnamefont{Pani}},
  \bibinfo{author}{\bibfnamefont{V.}~\bibnamefont{Cardoso}}, \bibnamefont{and}
  \bibinfo{author}{\bibfnamefont{T.}~\bibnamefont{Delsate}},
  \bibinfo{journal}{Phys. Rev. Lett.} \textbf{\bibinfo{volume}{107}},
  \bibinfo{pages}{031101} (\bibinfo{year}{2011}).

\bibitem[{\citenamefont{Avelino}(2012)}]{Avelino:2012ge}
\bibinfo{author}{\bibfnamefont{P.~P.} \bibnamefont{Avelino}},
  \bibinfo{journal}{Phys. Rev.} \textbf{\bibinfo{volume}{D85}},
  \bibinfo{pages}{104053} (\bibinfo{year}{2012}).

\bibitem[{\citenamefont{Pani et~al.}(2012)\citenamefont{Pani, Delsate, and
  Cardoso}}]{Pani:2012qb}
\bibinfo{author}{\bibfnamefont{P.}~\bibnamefont{Pani}},
  \bibinfo{author}{\bibfnamefont{T.}~\bibnamefont{Delsate}}, \bibnamefont{and}
  \bibinfo{author}{\bibfnamefont{V.}~\bibnamefont{Cardoso}}
  (\bibinfo{year}{2012}), \eprint{1201.2814}.

\bibitem[{\citenamefont{Deser and Gibbons}(1998)}]{Deser:1998rj}
\bibinfo{author}{\bibfnamefont{S.}~\bibnamefont{Deser}} \bibnamefont{and}
  \bibinfo{author}{\bibfnamefont{G.~W.} \bibnamefont{Gibbons}},
  \bibinfo{journal}{Class. Quant. Grav.} \textbf{\bibinfo{volume}{15}},
  \bibinfo{pages}{L35} (\bibinfo{year}{1998}).

\bibitem[{\citenamefont{Vollick}(2004)}]{Vollick:2003qp}
\bibinfo{author}{\bibfnamefont{D.~N.} \bibnamefont{Vollick}},
  \bibinfo{journal}{Phys. Rev.} \textbf{\bibinfo{volume}{D69}},
  \bibinfo{pages}{064030} (\bibinfo{year}{2004}).

\bibitem[{\citenamefont{Wohlfarth}(2004)}]{Wohlfarth:2003ss}
\bibinfo{author}{\bibfnamefont{M.~N.~R.} \bibnamefont{Wohlfarth}},
  \bibinfo{journal}{Class. Quant. Grav.} \textbf{\bibinfo{volume}{21}},
  \bibinfo{pages}{1927} (\bibinfo{year}{2004}).

\bibitem[{\citenamefont{Nieto}(2004)}]{Nieto:2004qj}
\bibinfo{author}{\bibfnamefont{J.}~\bibnamefont{Nieto}},
  \bibinfo{journal}{Phys. Rev.} \textbf{\bibinfo{volume}{D70}},
  \bibinfo{pages}{044042} (\bibinfo{year}{2004}).

\bibitem[{\citenamefont{Comelli}(2005)}]{Comelli:2005tn}
\bibinfo{author}{\bibfnamefont{D.}~\bibnamefont{Comelli}},
  \bibinfo{journal}{Phys. Rev.} \textbf{\bibinfo{volume}{D72}},
  \bibinfo{pages}{064018} (\bibinfo{year}{2005}).

\bibitem[{\citenamefont{Vollick}(2005)}]{Vollick:2005gc}
\bibinfo{author}{\bibfnamefont{D.~N.} \bibnamefont{Vollick}},
  \bibinfo{journal}{Phys. Rev.} \textbf{\bibinfo{volume}{D72}},
  \bibinfo{pages}{084026} (\bibinfo{year}{2005}).

\bibitem[{\citenamefont{Zinoviev}(2006)}]{Zinoviev:2006}
\bibinfo{author}{\bibfnamefont{Y.~M.} \bibnamefont{Zinoviev}},
  \bibinfo{journal}{Journal of High Energy Physics}
  \textbf{\bibinfo{volume}{2006}}, \bibinfo{pages}{009} (\bibinfo{year}{2006}).

\bibitem[{\citenamefont{Ferraro and Fiorini}(2008)}]{Ferraro:2008ey}
\bibinfo{author}{\bibfnamefont{R.}~\bibnamefont{Ferraro}} \bibnamefont{and}
  \bibinfo{author}{\bibfnamefont{F.}~\bibnamefont{Fiorini}},
  \bibinfo{journal}{Phys. Rev.} \textbf{\bibinfo{volume}{D78}},
  \bibinfo{pages}{124019} (\bibinfo{year}{2008}).

\bibitem[{\citenamefont{Fiorini and Ferraro}(2009)}]{Fiorini:2009ux}
\bibinfo{author}{\bibfnamefont{F.}~\bibnamefont{Fiorini}} \bibnamefont{and}
  \bibinfo{author}{\bibfnamefont{R.}~\bibnamefont{Ferraro}},
  \bibinfo{journal}{Int. J. Mod. Phys.} \textbf{\bibinfo{volume}{A24}},
  \bibinfo{pages}{1686} (\bibinfo{year}{2009}).

\bibitem[{\citenamefont{Ferraro and Fiorini}(2010)}]{Ferraro:2009zk}
\bibinfo{author}{\bibfnamefont{R.}~\bibnamefont{Ferraro}} \bibnamefont{and}
  \bibinfo{author}{\bibfnamefont{F.}~\bibnamefont{Fiorini}},
  \bibinfo{journal}{Phys. Lett.} \textbf{\bibinfo{volume}{B692}},
  \bibinfo{pages}{206} (\bibinfo{year}{2010}).

\bibitem[{\citenamefont{Gullu et~al.}(2010{\natexlab{a}})\citenamefont{Gullu,
  Sisman, and Tekin}}]{Gullu:2010pc}
\bibinfo{author}{\bibfnamefont{I.}~\bibnamefont{Gullu}},
  \bibinfo{author}{\bibfnamefont{T.~C.} \bibnamefont{Sisman}},
  \bibnamefont{and} \bibinfo{author}{\bibfnamefont{B.}~\bibnamefont{Tekin}},
  \bibinfo{journal}{Class. Quant. Grav.} \textbf{\bibinfo{volume}{27}},
  \bibinfo{pages}{162001} (\bibinfo{year}{2010}{\natexlab{a}}).

\bibitem[{\citenamefont{Alishahiha et~al.}(2010)\citenamefont{Alishahiha,
  Naseh, and Soltanpanahi}}]{Alishahiha:2010iq}
\bibinfo{author}{\bibfnamefont{M.}~\bibnamefont{Alishahiha}},
  \bibinfo{author}{\bibfnamefont{A.}~\bibnamefont{Naseh}}, \bibnamefont{and}
  \bibinfo{author}{\bibfnamefont{H.}~\bibnamefont{Soltanpanahi}},
  \bibinfo{journal}{Phys.Rev.} \textbf{\bibinfo{volume}{D82}},
  \bibinfo{pages}{024042} (\bibinfo{year}{2010}).

\bibitem[{\citenamefont{Gullu et~al.}(2010{\natexlab{b}})\citenamefont{Gullu,
  Sisman, and Tekin}}]{Gullu:2010em}
\bibinfo{author}{\bibfnamefont{I.}~\bibnamefont{Gullu}},
  \bibinfo{author}{\bibfnamefont{T.~C.} \bibnamefont{Sisman}},
  \bibnamefont{and} \bibinfo{author}{\bibfnamefont{B.}~\bibnamefont{Tekin}},
  \bibinfo{journal}{Phys. Rev.} \textbf{\bibinfo{volume}{D82}},
  \bibinfo{pages}{124023} (\bibinfo{year}{2010}{\natexlab{b}}).

\bibitem[{\citenamefont{De~Felice et~al.}(2012)\citenamefont{De~Felice,
  Gumjudpai, and Jhingan}}]{DeFelice:2012hq}
\bibinfo{author}{\bibfnamefont{A.}~\bibnamefont{De~Felice}},
  \bibinfo{author}{\bibfnamefont{B.}~\bibnamefont{Gumjudpai}},
  \bibnamefont{and} \bibinfo{author}{\bibfnamefont{S.}~\bibnamefont{Jhingan}}
  (\bibinfo{year}{2012}), \eprint{1205.1168}.

\bibitem[{\citenamefont{Clifton et~al.}(2012)\citenamefont{Clifton, Ferreira,
  Padilla, and Skordis}}]{Clifton:2011jh}
\bibinfo{author}{\bibfnamefont{T.}~\bibnamefont{Clifton}},
  \bibinfo{author}{\bibfnamefont{P.~G.} \bibnamefont{Ferreira}},
  \bibinfo{author}{\bibfnamefont{A.}~\bibnamefont{Padilla}}, \bibnamefont{and}
  \bibinfo{author}{\bibfnamefont{C.}~\bibnamefont{Skordis}},
  \bibinfo{journal}{Phys. Rept.} \textbf{\bibinfo{volume}{513}},
  \bibinfo{pages}{1} (\bibinfo{year}{2012}).

\bibitem[{\citenamefont{Casanellas et~al.}(2012)\citenamefont{Casanellas, Pani,
  Lopes, and Cardoso}}]{Casanellas:2011kf}
\bibinfo{author}{\bibfnamefont{J.}~\bibnamefont{Casanellas}},
  \bibinfo{author}{\bibfnamefont{P.}~\bibnamefont{Pani}},
  \bibinfo{author}{\bibfnamefont{I.}~\bibnamefont{Lopes}}, \bibnamefont{and}
  \bibinfo{author}{\bibfnamefont{V.}~\bibnamefont{Cardoso}},
  \bibinfo{journal}{Astrophys. J.} \textbf{\bibinfo{volume}{745}},
  \bibinfo{pages}{15} (\bibinfo{year}{2012}).

\bibitem[{\citenamefont{Delsate and Steinhoff}(2012)}]{Delsate:2012ky}
\bibinfo{author}{\bibfnamefont{T.}~\bibnamefont{Delsate}} \bibnamefont{and}
  \bibinfo{author}{\bibfnamefont{J.}~\bibnamefont{Steinhoff}}
  (\bibinfo{year}{2012}), \eprint{1201.4989}.

\end{thebibliography}

\end{document}